# Effect of doping on the thermoelectric properties of thallium tellurides using first principles calculations

Philippe Jund[a], Xiaoma Tao, Romain Viennois and Jean-Claude Tédenac

Institut Charles Gerhardt, Université Montpellier 2, Pl. E. Bataillon CC1503
34095 Montpellier, France
[a]pjund@univ-montp2.fr

**Keywords:** first-principles calculations, electronic structure, thermoelectricity, tellurides

Abstract.

We present a study of the electronic properties of $Tl_5Te_3$, $BiTl_9Te_6$ and $SbTl_9Te_6$ compounds by means of density functional theory based calculations. The optimized lattice constants of the compounds are in good agreement with the experimental data. The band gap of $BiTl_9Te_6$ and $SbTl_9Te_6$ compounds are found to be equal to 0.589 eV and 0.538 eV, respectively and are in agreement with the available experimental data. To compare the thermoelectric properties of the different compounds we calculate their thermopower using Mott's law and show, as expected experimentally, that the substituted tellurides have much better thermoelectric properties compared to the pure compound.

**Introduction**

Recently thermoelectric materials have attracted the interest of many researchers and a lot of investigations have been performed showing that high efficiencies could indeed be obtained [1]. Among the III-VI group, the Tl-Te system is of particular interest and has been the topic of many investigations for its thermoelectric properties [2-7]. This system exists in four crystalline phases: $Tl_2Te_3$, $TlTe$, $Tl_5Te_3$, and $Tl_2Te$ [8]. $Tl_5Te_3$ has a metallic behavior and exhibits a superconducting transition at low temperature [2,3,5]. The $Tl_5Te_3$ structure has four crystallographic sites: Tl1 site 16l; Tl2 site 4c; Te1 site 4a and Te2 site 8h. This is a $Cr_5B_3$-prototype structure with Pearson's symbol I4/mcm (140) [9]. According to B. Wölfing et al [10,11], the three Te atoms in $Tl_5Te_3$ are in the valence state $Te^{2-}$, therefore the five Tl atoms have to provide six electrons. This is achieved by expanding the formula unit to $Tl^{1+}_9Tl^{3+}Te^{2-}_6$, where the 4c site, which accommodates two Tl atoms per expanded formula unit, is equally occupied by $Tl^+$ and $Tl^{3+}$. Thus, the $Tl^{3+}$ can be substituted with the trivalent elements $Bi^{3+}$ and $Sb^{3+}$, resulting in the compounds $BiTl_9Te_6$ and $SbTl_9Te_6$. Some studies have already been performed on $BiTl_9Te_6$ [10-13] and $SbTl_9Te_6$ [11] due to their excellent thermoelectric properties and this explains why these compounds are the topic of the present study.

Nevertheless thallium is a toxic element which can not be used directly in thermoelectric materials. But a fundamental study of this material can be done to understand the properties of other kind of materials crystallizing in the same structure.

Even though the knowledge of the different physical properties of these compounds is important, there are only few studies focused on the electronic and mechanical properties of $Tl_5Te_3$, $BiTl_9Te_6$ and $SbTl_9Te_6$. The structural, elastic and thermodynamic properties obtained using first-principles calculations based on the Density Functional Theory (DFT) will be the topic of a separate publication [14]. In this work we focus on the calculated band structure of the different alloys and try to obtain the Seebeck coefficient at 300K using the Mott's law [15] in order to compare the potential thermoelectric properties of the different compounds.

The remainder of this paper is organized as follows. In Section 2, the method and the calculation details are described. In section 3, the electronic band structure and the Seebeck coefficients are presented and discussed. Finally, some conclusions are drawn in Section 4.

**Computational details and method**

First-principles calculations are performed by using the scalar relativistic all-electron Blöchl's projector augmented-wave (PAW) method [16,17] within the generalized gradient approximation (GGA), as implemented in the highly-efficient Vienna Ab initio Simulation Package (VASP) [18,19]. For the GGA exchange- correlation function, the Perdew-Wang parameterization (PW91) [20,21] is employed. Here we adopted the standard version of the PAW potentials for Tl, Te, Bi, and Sb atoms. A plane-wave energy cutoff of 300 eV is held constant for all the calculations. Brillouin zone integrations are performed using Monkhorst-Pack k-point meshes [22], with a k-point sampling of 9x9x9 for the $Cr_5B_3$-type structure which is the ground state structure of all the systems studied here [14]. The Methfessel-Paxton technique [23] with a smearing parameter of 0.2 eV is also used. The total energy is converged numerically to less than $1\times10^{-6}$ eV/unit. After structural optimization, calculated forces are converged to less than 0.01eV/Å.

The efficiency of thermoelectric generators determined by the dimensionless thermoelectric material figure of merit [1], $zT$:

$$zT = T\frac{S^2\sigma}{\kappa} \quad (1)$$

where $S$ is the thermoelectric power or Seebeck coefficient of the thermoelectric material, $\sigma$ and $\kappa$ are the electrical and thermal conductivities, respectively, and $T$ is the absolute temperature.

To determine the thermopower α, which is one of the variables entering zT, we have used Mott's law [15] which gives us:

$$\frac{\alpha}{T} = -\frac{\pi^2 k_B^2}{3e}\left(\frac{\partial \ln(\sigma(\varepsilon))}{\partial \varepsilon}\right)_{\varepsilon_F} \quad (1)$$

If we assume a negligible k-dependence of the group velocity close to the Fermi level and with the hypothesis of constant relaxation time, we obtain:

$$\frac{\alpha}{T} = -\frac{\pi^2 k_B^2}{3e}\left(\frac{\partial \ln(n(\varepsilon))}{\partial \varepsilon}\right)_{\varepsilon_F} \quad (2)$$

where n(ε) is the electronic density of states.
This is a crude approximation but as discussed by Tobola and coworkers [24], in favorable cases, eq. (2) can give qualitative information about the doping dependence of the thermopower.

**Results and discussion**

The total and partial density of states and band structures for the $Tl_5Te_3$, $BiTl_9Te_6$, and $SbTl_9Te_6$ compounds with the $Cr_5B_3$-prototype structure have been calculated and are shown in Fig.1, 2, and 3, respectively. Fig. 1(a) shows the calculated electronic density of states of $Tl_5Te_3$, while Fig. 1(b) shows the band structure of $Tl_5Te_3$ along the high symmetry directions Γ→X→P→N→Γ→M. There is a narrow band gap observed above the Fermi level, which is consistent with the metallic behaviour of $Tl_5Te_3$.

The electronic density of states for the $BiTl_9Te_6$ compound is shown in Fig. 2(a). The band gap of 0.589 eV at the Fermi level indicates that $BiTl_9Te_6$ is a semiconductor. The present calculated band gap is in agreement with the reported value $E_g \geq 0.4$ eV [10]. From Fig. 2(b) it can be seen that the valence band maximum occurs along the NΓ direction while the conduction band minimum occurs along the ΓX direction: $BiTl_9Te_6$ is thus an indirect band gap semiconductor.

The electronic density of states and band structure of $SbTl_9Te_6$ (not shown) are very similar to those of $BiTl_9Te_6$ and the predicted band gap of 0.538 eV is in good agreement with the estimated value of 0.5 eV [11]. The valence band maximum occurs along the NΓ direction while the conduction band minimum occurs along the ΓX direction: thus, similarly to $BiTl_9Te_6$, $SbTl_9Te_6$ is also an indirect band gap semiconductor.

To determine the thermopower at 300K of $Tl_5Te_3$ we simply applied Eq. (2) taking the logarithmic derivative of the electronic density of states at the Fermi level multiplied by $-2.441 \times 10^{-2}$ (keeping the energy in eV) and thus obtained a value of -5 μV/K which is very small and in relative good agreement with the experimental determination of 2 μV/K [25]. Nevertheless we have a disagreement concerning the sign of α. This is not surprising since looking at Fig.1a one sees that the Fermi level is located in a region in which the density of states changes very rapidly and thus a slight change in the approximations used in the calculations can induce a change of slope of the density of states at $E_f$ : we are probably at the limits of validity of the application of Mott's Law. For the determination of the thermopower of the doped compounds, we have to take into account that they are semiconducting and Mott's law is *a priori* only valid for metals. However, as can be seen in ref. 10, the resistivity has a metallic behaviour and the thermal dependence of the thermopower is relatively close to the linear behaviour expected in the Mott's law. Therefore applying the rigid band model [26], knowing the concentration of charge carriers and finally applying Mott's law, we can roughly estimate the thermopower. Since only one band is filled and because of the metallic behaviour observed in these compounds, we expect that this crude method can give qualitatively the thermoelectric properties of the doped compounds.

We can apply this method to the Bismuth doped compound since experimentally we know that at 300K the hole concentration in $BiTl_9Te_6$ is of $1.6 \; 10^{19}$ cm$^{-3}$ [10]. So we apply the rigid band model and lower the Fermi level on the DOS of Fig.2 until we reach the correct carrier concentration. Finally we apply Eq. (2) and obtain a thermopower at 300K of 300 µV/K for $BiTl_9Te_6$. This is extremely large but coherent with the experimental value of 260 µV/K found by Wolfing et al. [10]. In a similar way we can determine the thermopower of the Antimony doped compound knowing that the hole concentration at 300K is close to $1.5 \; 10^{20}$ cm$^{-3}$ [27]. We obtain a value of 66 µV/K which should be compared to the experimental value of 100 µV/K [27].

Of course this procedure needs to be checked on other compounds but even though it is a rough approximation it shows clearly that the Bi doped compound has much better thermoelectric properties compared to the pure Tl5Te3 alloy or the Sb doped compound, and this is in agreement with experimental results. Thus this procedure can be useful in the future to predict the thermoelectric properties of other compounds and give hints to the experimentalists.

**Conclusion**

We have presented results of first-principles calculations for the electronic structure of $Tl_5Te_3$, and $BiTl_9Te_6$, and $SbTl_9Te_6$ compounds. The results show that the pure compound is metallic while the Bi and Sb doped compounds are indirect semiconductors. In order to estimate the thermopower of these systems we have used Mott's law together with the rigid band model for the semiconductors. Even though this procedure is a very crude approximation of the true electronic behaviour as a function of temperature we have obtained a quantitative and qualitative agreement with the experimental results at 300K. In particular it permits to confirm the extremely high thermopower of $BiTl_9Te_6$ and the extremely small thermopower of Tl5Te3.

The present results give hints for the design of thermoelectric materials based on $Tl_5Te_3$ and should be used to stimulate future experimental and theoretical work.

**References**

[1] Snyder G.J. and Toberer E.S. Nature mater. 7, 105 (2008)

[2] A. Juodakis and C. R. Kannewurf, J. Appl. Phys. 39, 3003 (1968)

[3] E. Cruceanu, St. Sladaru, J. Mater. Sci. 4, 410 (1969)

[4] T. Ikari, K. Hashimoto, Phys. Stat. Sol. (b) 86, 239 (1978)

[5] J.D. Jensen, J.R. Burke, D.W. Ernst and R.S. Allgaier, Phys. Rev. B. 6, 319 (1972)

[6] K.J. Nordell, G.J. Miller, J. Alloys Compds. 241, 51 (1996)

[7] P.E. Lippens, L. Aldon, Solid State Commun. 108, 913 (1998)

[8] M.C. Record, Y. Feutelais and H.L. Lukas, Z. Metallkd. 88, 45 (1997)

[9] P. Villars, L.D. Calvert, *"Pearson's Handbook of Crystallographic Data for Intermetallic Phases"*, Vol. 1-4, ASM International, Materials Park (OH-USA), 1991

[10] B. Wölfing, C. Kloc, J. Teubner and E. Bucher, Phys. Rev. Lett. 86, 4350 (2001)

[11] B. Wölfing, Ph.D. thesis, Universitat Konstanz (2000)

[12] K. Kurosaki, A, Kosuga, S. Yamanaka, J. Alloys Compds. 351, 14 (2003)

[13] P.E. Blöchl, Phys. Rev. B 50, 17953 (1994)

[14] X. Tao, P. Jund and J.-C. Tédenac, Phy. Rev. B submitted.

[15] N.F. Mott and H. Jones: The *theory of the properties of metals and alloys* (Dover, New York, 1958), p.116.

[16] P.E. Blöchl, Phys. Rev. B 50, 17953 (1994)

[17] G. Kresse and D. Joubert, Phys. Rev. B 59, 1758 (1999)

[18] G. Kresse, J. Furthmuller, Phys. Rev. B 54, 11169 (1996)

[19] G. Kresse, J. Furthmuller, Comput. Mater. Sci. 6, 15 (1996)

[20] J.P. Perdew, Y. Wang, Phys. Rev. B 45, 13244 (1992)

[21] J.P. Perdew, J.A. Chevary, S.H. Vosko, K.A. Jackson, M.R. Pederson, D.J. Singh, C. Fiolhais, Phys. Rev. B 46, 6671 (1992)

[22] H.J. Monkhorst, J.D. Pack, Phys. Rev. B 13, 5188 (1976)

[23] M. Methfessel and A.T. Paxton, Phys. Rev. B 40, 3616 (1989)

[24] T. Stopa, J. Tobola, S. Kaprzyk, E. K. Hlil, D. Fruchart, J. Phys. : Cond. Matter. **18**, 6379 (2006)

[25] G.A. Gamal, M.M Abdalrahman, M.I Ashraf and H.J. Eman: Phys. Stat. Sol (a) Vol. 192 (2002), p. 322

[26] M.H. Cohen and V. Heine: Advan. Phys. Vol 7 (1958), p. 395

[27] J.T. Teubner, Ph.D. thesis, Universitat Konstanz (2001)

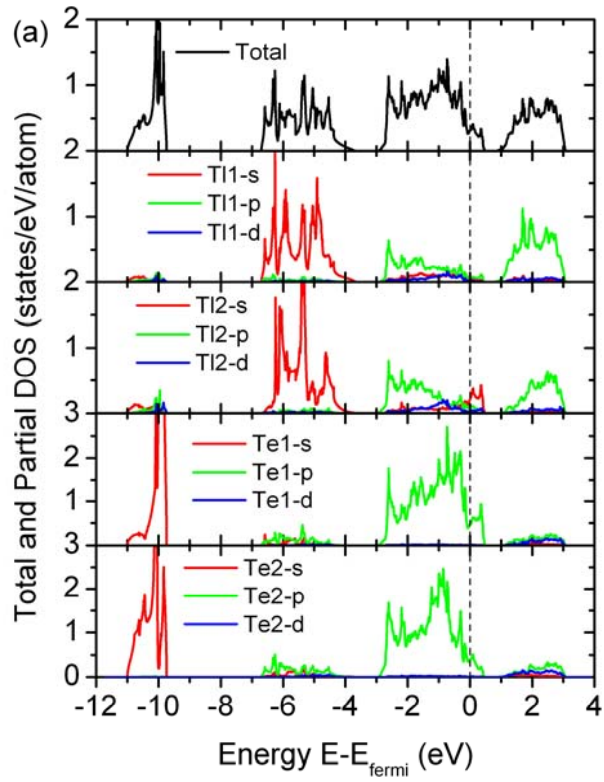

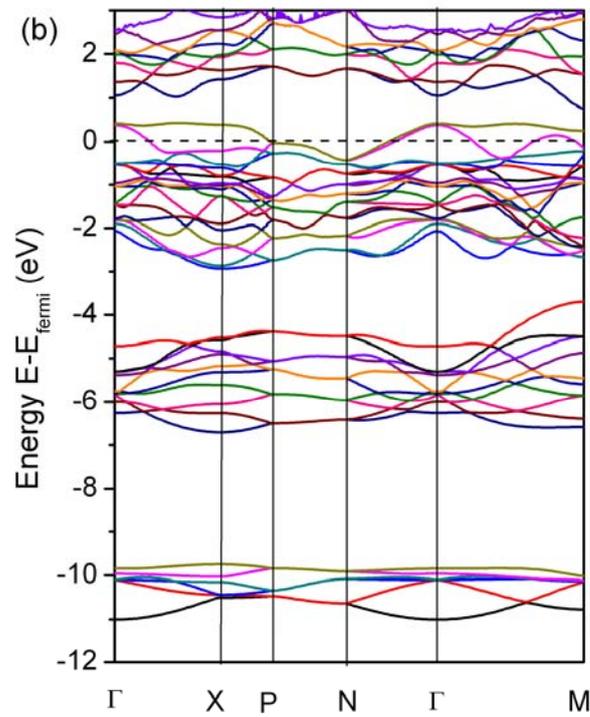

Fig.1

(a) Total and partial density of states of $Tl_5Te_3$; (b) band structure of $Tl_5Te_3$

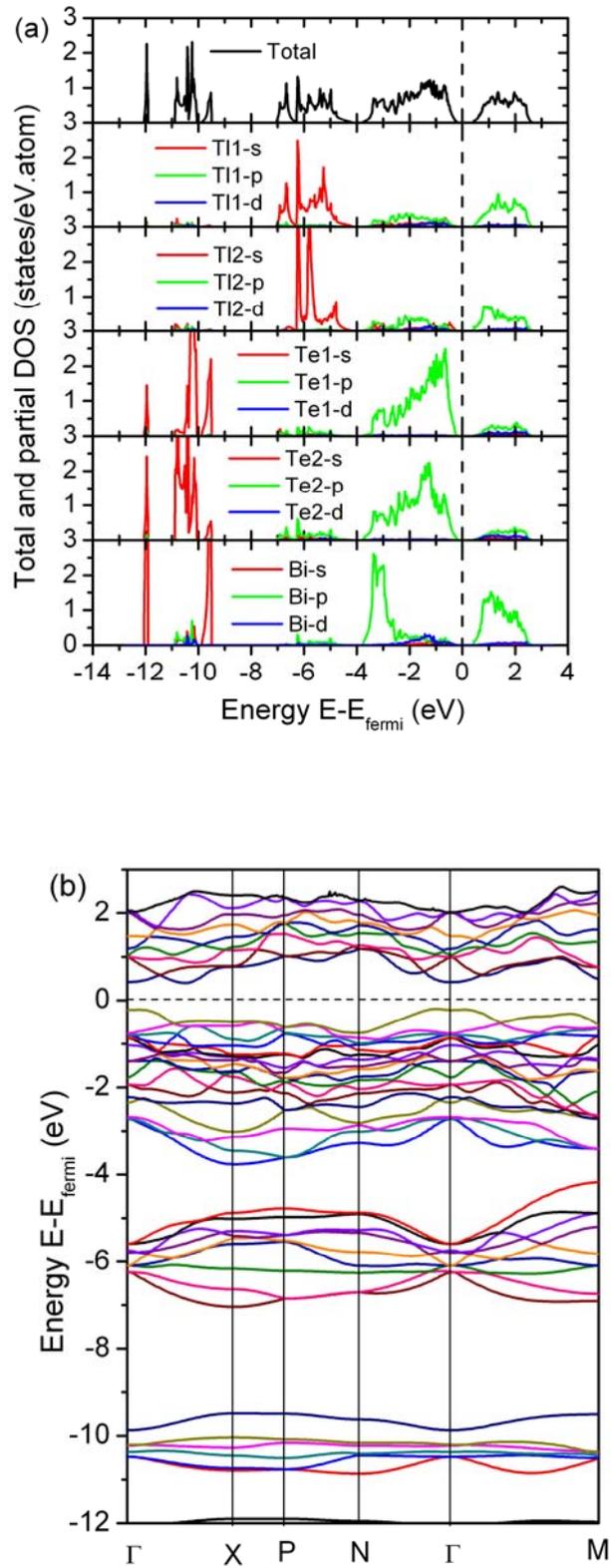

Fig. 2

(a) Total and partial density of states of $BiTl_9Te_6$; (b) band structure of $BiTl_9Te_6$